%% file: paper.tex
\documentclass[10pt,conference]{IEEEtran}
\IEEEoverridecommandlockouts
\usepackage{amsmath,amssymb,amsfonts}
\usepackage{algorithmic}
\usepackage{graphicx}
\usepackage{textcomp}
\usepackage{xcolor}

\usepackage[numbers,sort]{natbib}
\usepackage[colorlinks=false,hidelinks]{hyperref}
\usepackage{cleveref}
\usepackage[braket, qm]{qcircuit}
\usepackage{dblfloatfix}
\usepackage{csquotes}
\usepackage{paralist}

\def\BibTeX{{\rm B\kern-.05em{\sc i\kern-.025em b}\kern-.08em
    T\kern-.1667em\lower.7ex\hbox{E}\kern-.125emX}}
\begin{document}
\bstctlcite{IEEEexample:BSTcontrol}

\title{Warm-Starting the VQE with Approximate Complex Amplitude Encoding}

\author{\IEEEauthorblockN{
Felix Truger,
Johanna Barzen,
Frank Leymann, and
Julian Obst
}

\IEEEauthorblockA{
\small{\textit{Institute of Architecture of Application Systems,}
\textit{University of Stuttgart,}
\texttt{\{lastname\}@iaas.uni-stuttgart.de}}
}
}

\maketitle

\begin{abstract}
\input{Content/abstract}
\looseness=-1
\end{abstract}


\section{Introduction}
Quantum computers are expected to excel in tasks directly related to the properties of quantum-mechanical systems, e.g., in quantum chemistry and condensed matter physics,~\cite{feynman1982simulating,preskill2018quantum,Tilly2022VQEreview,peruzzo2014variational}.
The \textit{Variational Quantum Eigensolver (VQE)} is a hybrid quantum-classical algorithm to obtain the ground state energy of a given Hamiltonian operator describing such a system.
As a \textit{Variational Quantum Algorithm (VQA)}, the~VQE is suited for current \textit{Noisy Intermediate-Scale Quantum (NISQ)} devices with their limitations in the width and depth of executable quantum circuits~\cite{preskill2018quantum,Leymann2020_QuantumAlgorithmsBitterTurth}.
Therefore, the VQE makes use of shallow parameterized quantum circuits executed on NISQ hardware and optimization algorithms executed on classical computers.
This hybrid quantum-classical approach is deemed promising in the current era, where quantum computers are still error-prone and limited in number of qubits. 
Despite the bright prospects for the VQE, several potential obstacles remain~\cite{Tilly2022VQEreview}:
To evaluate the energy of a system, each optimization step of the VQE requires a multitude of measurements, that depends on the composition of the given Hamiltonian.
Moreover, the optimization landscape may be hard to navigate due to adverse effects such as barren plateaus and local optima, and the algorithm may exhibit insufficient convergence properties.
\looseness=-1

Warm-starts are known for their capability to mitigate some of these obstacles in quantum algorithms~\cite{Truger2023_warm_starting}.
Warm-starting techniques focus on the utilization of known or efficiently generated results instead of starting an algorithm from scratch, e.g., by making use of previous results or efficient approximation algorithms.
Several kinds of warm-starting techniques affect quantum algorithms in different ways.
For example, there are two major entry points for warm-starts of VQAs: Initial states and initial variational parameters.
On the one hand, such warm-starts can be realized through the encoding of prior knowledge into the initial state of a quantum circuit~\cite[cf.][]{egger2021warm,tate2023bridging}.
Thereby, the initial state is biased towards favorable solutions, as opposed to neutral initial states frequently assumed in conventional quantum algorithms.
On the other hand, warm-starts for VQAs can be realized by providing viable initial values for the parameters of the quantum circuit instead of a random parameter initialization, e.g., by transferring optimal parameter values from related problem instances or precomputing beneficial parameter initializations,~\cite[cf.][]{galda2021transferability,mitarai2022quadratic,sack2021quantum,shaydulin2023parameter}.

Encoding approximate eigenvectors via amplitude encoding yields a simple biased initial state to warm-start the VQE.
However, this warm-start is impractical due to the inefficiency of the encoding and current hardware limitations.
Moreover, it imposes certain restrictions on the Ansatz of the VQE, as we further elaborate in \Cref{sec:ws-vqe}.
In this work, we utilize \textit{Approximate Complex Amplitude Encoding (ACAE)}~\cite{mitsuda2022ACAE} to convert this idea into a warm-start via parameter initialization, that is more suitable for current hardware.
By means of a fidelity estimation from so-called classical shadows~\cite{huang2020predicting}, the variational ACAE algorithm provides an efficient approximate amplitude encoding of complex vectors into a quantum state.
We evaluate the performance benefits of the resulting warm-started VQE (henceforth WS-VQE) over the standard VQE.
The underlying warm-starting technique employed in WS-VQE can be used as a blueprint for warm-starting other VQAs.

The remainder of this paper is organized as follows:
In \Cref{sec:bg}, we introduce the background and fundamentals for this work.
\Cref{sec:rw} discusses related work and \Cref{sec:ws-vqe} introduces WS-VQE, which is evaluated and analyzed in \Cref{sec:eval}.
The results are further discussed in \Cref{sec:dis} and the paper is concluded with a summary and outlook in \Cref{sec:summary}.
\looseness=-1

\section{Background and Fundamentals}
\label{sec:bg}
This section introduces the VQE and ACAE in more detail to provide the background and fundamentals for WS-VQE.
\looseness=-1
\subsection{Variational Quantum Eigensolver}
The VQE follows the general construction of VQAs, i.e., it is a hybrid quantum-classical algorithm based on a parameterized quantum circuit, frequently referred to as \textit{Ansatz}, and a classical optimizer~\cite{peruzzo2014variational,cerezo2021variational}.
The purpose of the Ansatz is to prepare a quantum state representing a solution to the problem at hand.
The classical optimizer iteratively adjusts the parameter values for the Ansatz, that is in turn executed on a quantum device to assess the current solution and navigate to an optimum.
\looseness=-1

\citet{peruzzo2014variational} introduced the VQE to determine eigenvalues of operators more efficiently than possible with Quantum Phase Estimation.
At its heart, the VQE makes use of \textit{Quantum Expectation Estimation (QEE)}, a subroutine that evaluates the expectation value of the given Hamiltonian for the state prepared by the Ansatz.
For the measurements on the quantum computer, the Hamiltonian is decomposed into a real-valued linear combination of tensor products of the identity $I$ and the Pauli operators $X$, $Y$, and $Z$.
The overall expectation value of the Hamiltonian $H$ is computed as per \Cref{eq:expectation}, i.e., as the weighted sum of the expectation values of each of the Pauli strings $P_i$ in the linear combination, where $x_i$ are the real-valued coefficients of $H$'s Pauli decomposition.
\begin{equation}
\label{eq:expectation}
\langle H\rangle = \sum_{i} x_i\langle P_i \rangle
\end{equation}
Therefore, the state prepared by the Ansatz needs to be measured multiple times for different measurement bases as prescribed by the Pauli decomposition.
These measurements require the preparation of measurement circuits with appropriate rotations to adjust the measurement basis for each qubit.
However, there are various strategies of grouping Pauli strings that can be measured jointly, which can reduce the number of measurement circuits significantly~\cite{Tilly2022VQEreview}.
Nonetheless, the QEE subroutine typically requires a multitude of measurement circuits to cover all Pauli strings composing $H$.
To obtain results of a certain precision, each measurement needs to be repeated $N_\text{shots}$ times.
Thus, the total number of shots required for each call of the QEE subroutine amounts to $N_\text{shots}\times n_\text{meas.circuits}$, where $n_\text{meas.circuits}$ is the number of measurement circuits, i.e., the number of groups of jointly measurable Pauli strings.\looseness=-1

The variational optimization of the VQE's Ansatz parameters aims to prepare a state that minimizes $H$'s expectation value.
Optimization typically starts from a random initial parameterization or beneficial values determined through specific methods~\cite{Tilly2022VQEreview}.
Thereby, the VQE eventually prepares an approximate ground state of $H$ with the lowest expectation.
Potential applications include determining electronic ground state energies and molecular potential energy surfaces in quantum chemistry~\cite{Li2019vqeQuantumChemistry}, strongly correlated systems in condensed matter physics~\cite{HeadMarsden2021quantumInfomration}, protein folding for drug discovery~\cite{Mustafa2022drugDiscovery,barkoutsos2021quantum}, material design~\cite{barkoutsos2021quantum}, and chemical engineering~\cite{bernal2022perspectives}.
The remainder of this work focuses on the underlying mathematical problem tackled with VQE, namely to determine the lowest eigenvalue and corresponding eigenvector of Hermitian matrices.

\subsection{Approximate Complex Amplitude Encoding}
\label{subsec:ACAE}
Amplitude encoding refers to encoding a (complex) vector \mbox{$\vec{\mathbf{x}} \in \mathbb{C}^n$} with $\vec{\mathbf{x}} = (x_0, \dots, x_{n-1})$ of unit length into the amplitudes of a quantum state $\ket{\vec{\mathbf{x}}}$, as shown in~\Cref{eq:ampl-encoding}~\cite{Weigold2020_DataEncodingPatterns,schuld2018supervised}.
\begin{equation}
\label{eq:ampl-encoding}
    \ket{\vec{\mathbf{x}}} = \sum_{i=0}^{n-1} x_i \ket{i}
\end{equation}

The main advantage of this encoding is, that storing data of length $n$ requires only $\log n$ qubits.
However, exact amplitude encoding is infeasible because a circuit depth of at least $\frac{1}{n}2^n$ is required for the preparation of an arbitrary state on $n$ qubits~\cite{schuld2018supervised,Shende2006Synthesis}.
\textit{Approximate amplitude encoding}~\cite{nakaji2022AAE} suggests a VQA that approximately encodes vectors into the amplitudes of a quantum state by training a shallow Ansatz.
However, this approach only supports the encoding of real-valued data.

Recently,~\citet{mitsuda2022ACAE} proposed \textit{Approximate Complex Amplitude Encoding (ACAE)} based on classical shadows~\cite{huang2020predicting}.
A classical shadow is an approximate classical description of a quantum state that is generated with few measurements.
It can be used to estimate different properties of the captured quantum state.
In the case of ACAE, classical shadows are employed to estimate the fidelity between a model state with the density operator $\rho_\mathrm{model}(\theta)$ and a target state with the density operator $\rho_\mathrm{target}$.
Based on the estimated fidelity, an Ansatz can be variationally optimized to approximately encode the target state described by $\rho_\mathrm{target}$.
Classical shadows are created by applying random unitary transformations to the state before measurements.
For the fidelity estimations in ACAE, the random unitaries are taken from the n-qubit Clifford group.
Results of multiple shots with different unitaries are averaged to obtain an expectation value.
The general idea is to classically undo these steps for each measurement result $\hat{b}_i$. 
The averaging operation is viewed as a quantum channel $\mathcal{M}$, that is reverted by the inverse $\mathcal{M}^{-1}$.
To undo the random Clifford unitary $U$, the inverse transformation $U^\dagger$ is applied.
The result $\hat{\rho}_i$ of undoing these operations is called a snapshot:
\looseness=-1

\begin{equation}
\hat{\rho}_i = \mathcal{M}^{-1} \left(U_i^\dagger |\hat{b}_i\rangle\langle\hat{b}_i| U_i\right)    
\end{equation}

A classical shadow of the original state is a collection of $N_\text{snaps}$ snapshots.
From this approximate classical description of the state, we can estimate certain properties.
For instance, the expectation value of an observable.
Since the expectation value of an observable $O$ in a state with the density operator $\rho$ is $\langle O\rangle=\operatorname{Tr}(O \rho)$, it is estimated by $\langle \hat{O} \rangle$ in \Cref{eq:expval}.
\looseness=-1
\begin{equation}
\label{eq:expval}
\langle \hat{O} \rangle = \frac{1}{N_\mathrm{snaps}} \sum_{i=1}^{N_\mathrm{snaps}} \operatorname{Tr}(O \hat{\rho}_i)    
\end{equation}

Setting $O = \rho_\mathrm{target}$ allows us to estimate the fidelity in ACAE.
As shown in the supplementary material provided by \citet{huang2020predicting}, $\mathcal{M}^{-1}$ boils down to \Cref{eq:inverted-channel}.
\begin{equation}
\label{eq:inverted-channel}
    \mathcal{M}^{-1}(\rho) = (2^n + 1) \rho - I
\end{equation}

Thus, the fidelity can be estimated by $\hat{f}(\theta)$:
\begin{equation}
\label{eq:fidelity}
\begin{split}
    \hat{f}(\theta) &= \frac{1}{N_\mathrm{snaps}} \sum_{i=1}^{N_\mathrm{snaps}} \operatorname{Tr}(\rho_\mathrm{target} \hat{\rho}_i)\\
    &= \frac{1}{N_\mathrm{snaps}} \sum_{i=1}^{N_\mathrm{snaps}} (2^n + 1) \langle \hat{b}_i | U_i \rho_\mathrm{target} U_i^\dagger | \hat{b}_i \rangle - 1
\end{split}
\end{equation}

\citet{mitsuda2022ACAE} used ACAE to perform amplitude encoding of test and training data for a quantum classifier.
\looseness=-1

\begin{figure*}[b]
    \centering
    \includegraphics[width=0.68\linewidth,trim=0 65 370 0, clip]{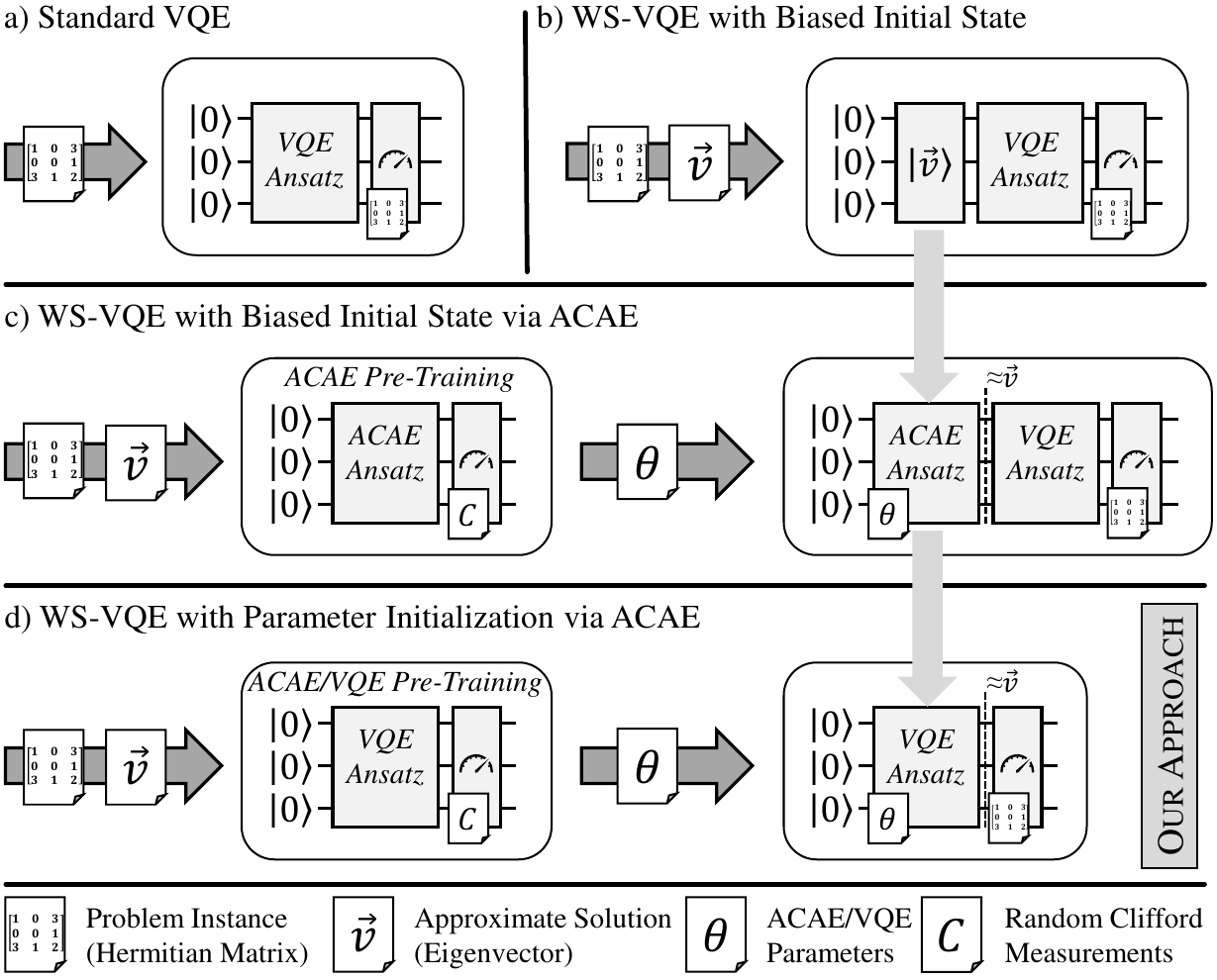}
    \caption{Shaping WS-VQE: 
    \textbf{a)} The standard VQE without warm-start.
    \textbf{b)} A simple warm-start for the VQE using amplitude encoding to prepend a biased initial state based on an approximate eigenvector $\vec{v}$ to the VQE Ansatz.
    \textbf{c)} A warm-start for the VQE that prepends a pretrained ACAE Ansatz instead of amplitude encoding.
    \textbf{d)} Our approach, a warm-start for VQE based on ACAE pretraining of the VQE Ansatz to approximately encode a given approximate eigenvector.
    \looseness=-1}
    \label{fig:WS-VQE-evolution}
\end{figure*}

\section{Related Work}
\label{sec:rw}
\citet{Tilly2022VQEreview} provide a comprehensive review of the VQE as well as methods and best practices related to it, however, not explicitly focusing on warm-starts for the VQE.
In previous work, we have conducted a mapping study for research on warm-starting techniques in the quantum computing domain, including warm-starts applicable to the VQE~\cite{Truger2023_warm_starting}:
\citet{Zhang2021adaptiveAnsatzConstruction} propose an adaptive construction of Ansatz circuits that takes information obtained from a classical approximation into account.
\citet{grimsley2023adaptive} propose dynamically growing the Ansatz during the execution of the VQE and recycling previous parameterizations.
Moreover, the VQE is compatible with meta-learning, i.e., classical machine learning models trained to take over the task of the optimizer in VQAs~\cite{verdon2019meta,wilson2021meta}, and plugging together multiple optimization steps that utilize optimized parameters from one step to initialize the next~\cite{tao2023laws}.
Most relevant to our work, various techniques concerning the parameter initialization of the VQE have been proposed.
Machine learning methods enable generating circuits and viable initial parameter values for each problem instance, that can be further optimized~\cite{dborin2022matrix,rudolph2022synergy}.
Exploiting the classically feasible simulation of Clifford circuits can yield viable parameter initializations for the VQE~\cite{Ravi2022CAFQA,mitarai2022quadratic}.
Other approaches try to obtain viable parameter initializations for a parameterized Hamiltonian, thus taking advantage of the continuity of the problem to obtain initializations for any parameterization of the Hamiltonian~\cite{cervera2021meta,Harwood2022VAQCVQE}.
Similarly, the VQE has been shown to be compatible with parameter transfers, where optimized parameters from one instance are reused to initialize the algorithm for a similar problem instance~\cite{skogh2023accelerating,kanno2021many}.

Apart from these concrete techniques, our \textit{Quokka} ecosystem, which enables the utilization of workflow technology for the service-based execution of VQAs, includes a \textit{warm-starting service}, that facilitates the precomputation of parameter initializations and biased initial states in quantum workflows~\cite{Beisel2023_Quokka}.
Our approach for WS-VQE in this work is different from the aforementioned warm-starts in that it is compatible with any given approximation of the problem and Ansatz for the VQE.
Moreover, since it does not require changes to the VQE, it is also compatible with adaptations of the VQE, including some of the aforementioned methods.

Furthermore, the VQE's measurements could be implemented based on classical shadows~\cite{Tilly2022VQEreview}.
A derandomized variant of classical shadows has shown potential in experiments~\cite{huang2021derandomizedCS}.
However, the performance of classical shadows compared to the grouping of Pauli strings and other grouping methods is still unclear~\cite{Tilly2022VQEreview}.
In any case, WS-VQE is also compatible with VQE implementations based on classical shadows.
\looseness=-1

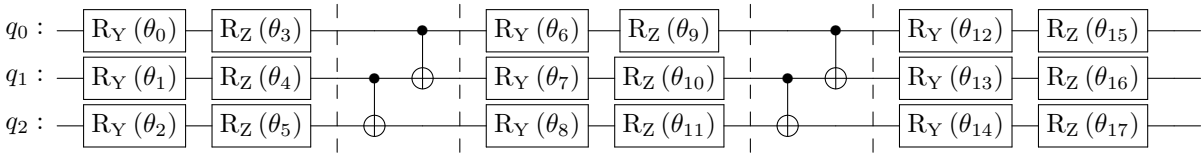
\begin{figure*}[b]
    \centering
    \input{Figures/circuit_subscripts}
    \vspace{3mm}
    \caption{Qiskit's hardware efficient SU(2) Ansatz (\texttt{EfficientSU2}) with two repetitions as used in the evaluation~\cite{qiskit}.}
    \label{fig:circuit}
\end{figure*}

\section{Warm-Starting the VQE}
\label{sec:ws-vqe}
In this section, we introduce our approach for warm-started VQE with ACAE (WS-VQE).
First, we briefly discuss a simple warm-start for the VQE via a biased initial state.
Then, we explain how ACAE helps in converting the impractical biased initial state into a practically usable parameter initialization for the VQE, hence shaping WS-VQE as illustrated in \Cref{fig:WS-VQE-evolution}.

\subsection{Biased Initial State}
\Cref{fig:WS-VQE-evolution}a) depicts the circuit of the standard VQE consisting of the Ansatz and measurements that depend on the problem instance, a Hermitian $H$.
Given an approximation $\vec{v}$ for an eigenvector of $H$ that corresponds to $H$'s lowest eigenvalue, a simple warm-start for VQE consists of prepending the amplitude encoding of $\vec{v}$ to the Ansatz as shown in \Cref{fig:WS-VQE-evolution}b) and starting the optimization with the variational parameters initialized to $\vec{0}$.
Intuitively, this warm-start can be viable since the optimized Ansatz of VQE encodes the eigenvector corresponding to the lowest eigenvalue of $H$.
Therefore, the encoded approximate eigenvector $\vec{v}$ and associated solution quality are trivially retained as long as the VQE's Ansatz parameterization results in the identity.
In turn, the solution can be improved from this state by optimizing the variational parameters.
However, the VQE Ansatz may therefore only contain gates that cancel out through neutralizing parameter values, typically $0$, or cancel out each other, such as CNOT gates uncomputing each other.
But also other parameter initializations evaluating to the identity are conceivable~\cite{grant2019initialization}.

Amplitude encoding is impractical due to unfavorable scaling, as detailed in \Cref{subsec:ACAE}.
As shown in \Cref{fig:WS-VQE-evolution}c), approximately encoding $\vec{v}$ with ACAE yields an alternative biased initial state.
However, the restrictions for the VQE Ansatz remain when a pretrained ACAE Ansatz is prepended to it.
\looseness=-1

\subsection{Parameter Initialization}
The restrictions for the VQE's Ansatz mentioned above can be circumvented by utilizing ACAE's approximate encoding to obtain a parameter initialization for the VQE instead of a biased initial state, which is illustrated in \Cref{fig:WS-VQE-evolution}d).
Since both the VQE and ACAE are VQAs with a relatively free choice for the Ansatz (see~\Cref{sec:bg}), the VQE's Ansatz can be pretrained using ACAE to encode $\vec{v}$ and thereby obtain initial parameters for the VQE.
Hence, the main idea of WS-VQE is to use the same Ansatz for both the VQE and ACAE, and conduct a two-step optimization of the Ansatz:
First, ACAE starts with a random parameter initialization to determine parameters for the Ansatz that approximately encode $\vec{v}$.
Then, the VQE is initialized with the parameters obtained for the encoding of $\vec{v}$ and continues optimizing the Ansatz to improve the solution.
The main advantage of using ACAE instead of starting VQE training from scratch is that ACAE can be expected to require fewer measurements.
Recall, that the VQE requires multiple measurement circuits for different Pauli strings to estimate the expectation value of $H$, each measurement circuit consuming $N_\text{shots}$ shots in every iteration, whereas ACAE utilizes classical shadows consuming $N_\text{snaps}$ single-shot snapshots to estimate the fidelity.
Thus, depending on $H$'s composition, pretraining the Ansatz with ACAE may be significantly cheaper than training the VQE directly in terms of overall shots.

\section{Evaluation}
\label{sec:eval}
This section first provides all necessary details on the setup and configuration of the evaluation before comparing the results obtained for both VQE and WS-VQE.
Afterward, optimization landscapes of VQE and ACAE are analyzed to provide deeper insights into WS-VQE's two-step optimization.\looseness=-1
\subsection{Setup and Configurations}
\subsubsection{Problem Instances}
\label{subsec:problem-instances}
For the evaluation of our approach, we generated $500$ problem instances to examine different VQE and WS-VQE configurations.
Each problem instance is a randomly generated $8\times8$ Hermitian matrix. 
The matrices are sparse, with $50\%$ probability for each entry of being either zero or a random complex number in the interval $[-5, 5]+[-5, 5]i$.
Reference eigenvalues for these instances were computed using the \texttt{NumPyMinimumEigensolver} provided by Qiskit~\cite{qiskit}.
Thus, approximation ratios can be computed as $r_\text{appr}=\frac{\lambda}{\lambda_{\text{ref}}}$, where $\lambda$ is an eigenvalue computed by the VQE and $\lambda_\text{ref}$ is the classically computed minimum eigenvalue of the instance.
Hence, $r_\text{appr}=1$ corresponds to an optimal solution.
\looseness=-1

\subsubsection{Classical Approximation}
For WS-VQE, we utilize a simple classical approximation of an eigenvector corresponding to the lowest eigenvalue based on the power method~\cite{quarteroni2006numerical}.
First, we apply the Gershgorin circle theorem~\cite{gershgorin1931circles} to determine a lower bound for the eigenvalues of $H$.
The theorem implies that each eigenvalue of $H$ is located within either of the circles surrounding $H_{ii}$ with a radius of $R_i=\sum_{k \neq i}|H_{ik}|$.
Based on the lower bound $\mu = \underset{i}{\operatorname{min}}\{H_{ii} - R_i\}$, the inverse power method allows us to obtain an approximation of an eigenvector to the eigenvalue closest to $\mu$, i.e., an approximate eigenvector to the lowest eigenvalue of $H$.
Starting with a random initial vector $q_{(0)}$, approximating the eigenvector boils down to \Cref{eq:approx-eigenv}.
We provide $q_{(3)}$ as approximate eigenvector to WS-VQE.
\looseness=-1
\begin{equation}
\label{eq:approx-eigenv}
\begin{split}
M&=(A-\mu I)^{-1}\\
z_{(k)} &= Mq_{(k-1)}\\
q_{(k)} &= \frac{z_{(k)}}{{||{z_{(k)}}||}_2}
\end{split}
\end{equation}

\subsubsection{Ansatz and VQE Implementation}
We employ Qiskit's hardware efficient SU(2) Ansatz (\texttt{EfficientSU2}), which is a multipurpose Ansatz that has been shown to be sufficiently expressive to prepare arbitrary quantum states~\cite{funcke2021dimensional}.
It consists of alternating rotation and entanglement blocks as depicted in~\Cref{fig:circuit}.
Since we consider $8\times8$ matrices as problem instances, the quantum circuit operates on $\log_2(8)=3$ qubits.
Particularly, we choose to generate the Ansatz with two repetitions of the alternating layers and a final rotation layer, resulting in a total of 18 variational parameters.
Two repetitions have been shown to provide a maximally expressive Ansatz for 3 qubits~\cite{funcke2021dimensional}.
Moreover, we utilize the VQE implementation and a noiseless quantum simulator provided by Qiskit for all executions of the VQE.
\looseness=-1

\subsubsection{Shots and Snapshots}
\label{subsec:shots}
\begin{figure}[h]
    \centering
    \includegraphics[trim=0 10 0 8, clip]{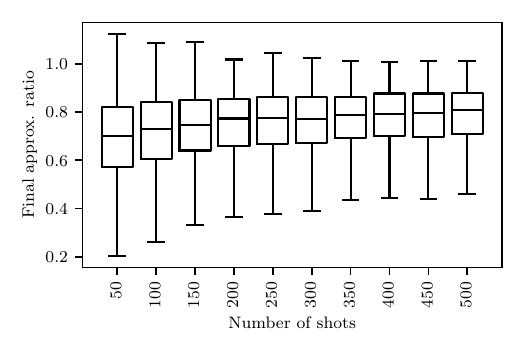}
    \caption{Median approximation ratio (horizontal markers) reached after $80$ iterations of VQE on $1\,000$ random problem instances per number of shots.}
    \label{fig:vqe_shots}
\end{figure}

From preliminary experiments summarized in~\Cref{fig:vqe_shots} we estimated that $N_\text{shots}=200$ in VQE's circuit executions yields a reasonable cost-benefit ratio.
Moreover, we take $N_\text{snaps}=400$ snapshots, i.e., $N_\text{snaps}$ single shots, for the fidelity estimation in ACAE.
Thereby, we ensure that ACAE's fidelity estimations are sufficiently accurate. 
For simplicity, we reduce the classical effort of the random Clifford measurements in ACAE by using the same $N_\text{snaps}$ Clifford unitaries throughout the optimization instead of generating new random Clifford unitaries in each iteration.
The Clifford operators are sampled uniformly at random using Qiskit's implementation of~\citeauthor{Bravyi2021Clifford}'s method~\cite{Bravyi2021Clifford}.
\looseness=-1

As mentioned above, computing the expectation value in the VQE consumes $N_\text{shots}$ shots for multiple measurement circuits, whereas ACAE consumes only $N_\text{snaps}$ shots to estimate the fidelity.
For the presentation of our evaluation results for WS-VQE, we prepend ACAE iterations of the pretraining to the subsequent VQE iterations.
To account for the difference in quantum computational effort, ACAE iterations are rescaled to an equivalent of VQE iterations with at least the same total number of shots. 
For instance, assuming one function evaluation per iteration, $20$ iterations of ACAE pretraining consume $400$ shots each, i.e., $8\,000$ shots in total and would be counted as an equivalent of $3$ VQE iterations, where $15$ measurement circuits consume $200$ shots each, i.e., $9\,000$ shots in total.
\looseness=-1

\subsubsection{Parameter Initialization}
All initial ACAE and VQE parameter values are drawn uniformly at random from $[-\pi,\pi]$.

\subsubsection{Optimizer Configuration}
\label{subsec:optimizer-config}
We employ the gradient-free~\texttt{COBYLA}~\cite{cobyla} as the classical optimizer in both ACAE and VQE.
\texttt{COBYLA} is known to perform reasonably well in noise-free simulation of VQAs~\cite{pellow2021comparison}.
For the pretraining of ACAE, we allow up to $50$ iterations, whereas we execute up to $100$ iterations for each run of the VQE.
Moreover, we set the initial step size parameter~\texttt{rhobeg} $(\varrho)$ of \texttt{COBYLA} to $\varrho_\text{ACAE}=\frac{1}{4}\pi$ for ACAE and $\varrho_\text{VQE}=\frac{3}{8}\pi$ for the VQE following our observations in preliminary experiments.
For WS-VQE, we assume that a reduction of \texttt{rhobeg} is beneficial due to the head start provided by ACAE pretraining.
Therefore, we expect WS-VQE to start closer to an optimum and require smaller changes to the parameters.
To confirm this assumption, we run WS-VQE with both $\varrho_\text{VQE}$ and $\varrho_\text{WS-VQE}^\text{static}=\frac{1}{2} \cdot \varrho_\text{VQE}$.
In addition, we consider a flexible setting where \texttt{rhobeg} depends on the final estimated fidelity $f_\text{final}$ achieved during ACAE pretraining.
A fidelity close to $1$ indicates a successful encoding of the approximated eigenvector, and therefore smaller steps should be required during the subsequent execution of the VQE.
In contrast, if the fidelity is close to $0$, the pretraining is considered as less successful and therefore more leeway should be given to the optimizer, since it may start farther from a global optimum.
This is taken into account by setting $\varrho_\text{WS-VQE}^\text{dynamic}=\frac{1}{f_\text{final}} \cdot \frac{1}{4} \cdot \varrho_\text{VQE}$.
Therefore, $\varrho_\text{WS-VQE}^\text{dynamic}$ ranges from $\frac{1}{4} \cdot \varrho_\text{VQE}$ in case that $f_\text{final}$ is $1$, i.e., perfect ACAE, over $\frac{1}{2} \cdot \varrho_\text{VQE}(=\varrho_\text{WS-VQE}^\text{static})$ in the case $f_\text{final}=0.5$ and $\varrho_\text{VQE}$ for $f_\text{final}=0.25$ to virtually $\infty$ in the worst case.

\subsubsection{Reproducibility}
The WS-VQE implementation used for the evaluation as well as all detailed results can be found in the GitHub repository associated with this work~\cite{gitRepo}.

\subsection{Results}
\label{sec:results}
\begin{figure}[b]
    \centering
    \includegraphics[trim=0 10 0 8, clip]{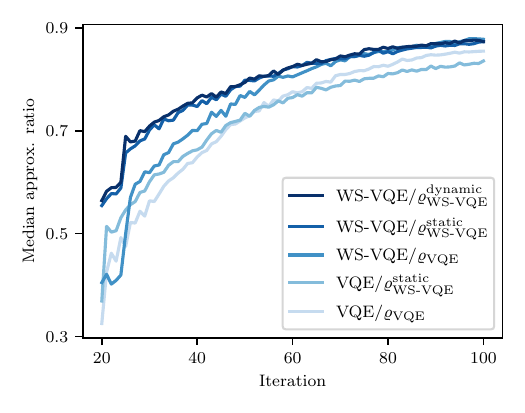}
    \caption{Progress of the (WS-)VQE optimization: Median approximation ratio after each iteration $i\in\{20, \dots, 100\}$ for the $500$ problem instances (see \Cref{subsec:problem-instances}) with different initial step sizes $\varrho$, as declared in \Cref{subsec:optimizer-config}.}
    \label{fig:results_vqe_comparison}
\end{figure}
\begin{figure*}[b]
    \includegraphics[trim=0 9 0 2, clip]{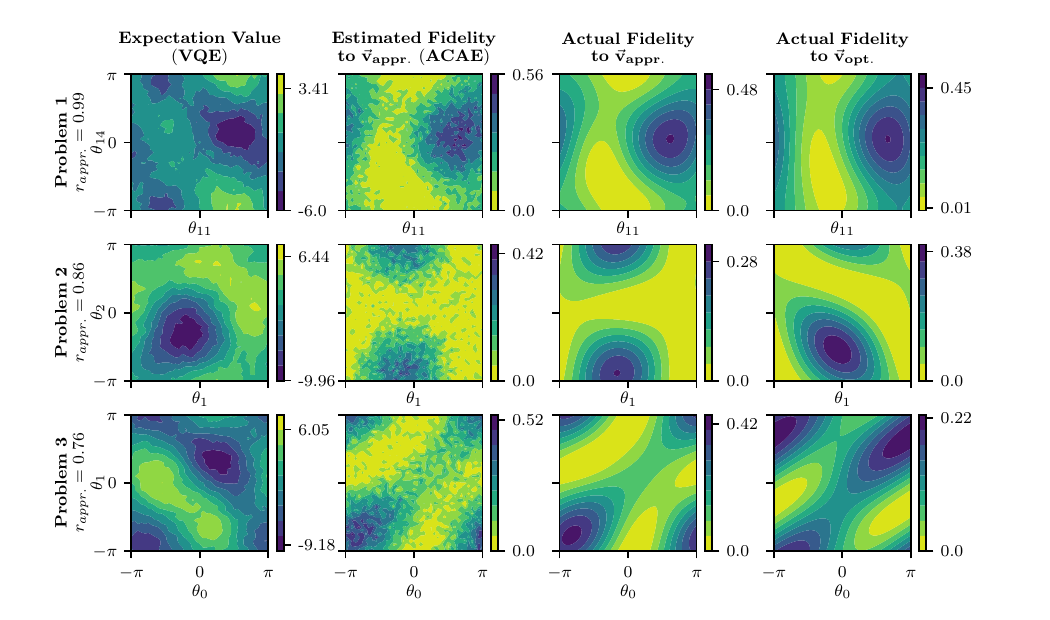}
    \caption{Slices of the parameter space of VQE and ACAE (left) for three different problem instances as compared to the actual fidelity of the quantum states prepared by the Ansatz to the approximate and an optimal eigenvector of the instance, $\vec{v}_\text{approx.}$ and $\vec{v}_\text{opt.}$, (right). Each line corresponds to a problem instance as described in \Cref{subsec:problem-instances} for which a random two-dimensional slice of the full 18-dimensional parameter space of the Ansatz depicted in \Cref{fig:circuit} is plotted.
    The remaining parameter values were selected uniformly at random from $[-\pi, \pi]$. Labels on the color bars indicate minimum and maximum values observed.\looseness=-1}
    \label{fig:optimization_landscapes}
\end{figure*}
\Cref{fig:results_vqe_comparison} shows the progress of the optimization of the VQE and WS-VQE with different optimizer configurations as stipulated above.
Evidently, the utilization of ACAE for the parameter initialization in WS-VQE leads to significant improvements of the median approximation ratio over standard VQE.
While the median approximation ratio reached after $20$ iterations is situated just above $0.3$ for the VQE, the WS-VQE variant can reach a median approximation ratio well above $0.5$, depending on the optimizer configuration.
The final median approximation ratios obtained with WS-VQE are also above those of the standard VQE.
Moreover, the different optimizer configurations for WS-VQE discussed above significantly impact the performance of WS-VQE.
As shown in \Cref{fig:results_vqe_comparison}, WS-VQE optimized with $\varrho_\text{VQE}$ reaches lower median approximation ratios than the standard VQE for a portion of the early iterations.
In contrast, WS-VQE/$\varrho_\text{WS-VQE}^\text{static}$, where \texttt{COBYLA}'s \texttt{rhobeg} parameter is halved, performs significantly better up until just above $65$ iterations where it starts to overlap with WS-VQE/$\varrho_\text{VQE}$.
The median approximation ratios obtained with WS-VQE/$\varrho_\text{WS-VQE}^\text{dynamic}$ are above those of all other variants until they also start to overlap with WS-VQE/$\varrho_\text{WS-VQE}^\text{static}$ in the 50th iteration.
These results confirm our assumptions regarding the impact of different optimizer settings for WS-VQE. 
However, these differences appear to fade as the optimization progresses.
Moreover, the results of VQE/$\varrho_\text{WS-VQE}^\text{static}$, which were included for comparison purposes, show that the advantages of WS-VQE are not merely the impact of the different optimizer setting.
\looseness=-1

\vspace{-5mm}
\subsection{Analysis of Optimization Landscapes}
\label{sec:landscapes}
Next, we further analyze the two-step optimization of \mbox{WS-VQE} by illustrating parts of the parameter space for the variational optimization.
The optimization landscapes plotted in \Cref{fig:optimization_landscapes} show expectation values and fidelities in a random two-dimensional slice of the parameter space of the Ansatz depicted in \Cref{fig:circuit}.
Values for parameters not annotated on the axes were selected uniformly at random for each problem instance.
Expectation values and fidelities, respectively, were evaluated for each point of an equidistant grid in the space $[-\pi, \pi]\times[-\pi, \pi]$ with a distance of $\frac{\pi}{20}$.
The problem instances and approximate eigenvectors are identical to the first three instances from the experiments above (cf. \Cref{fig:results_vqe_comparison}).
The approximation ratios are $0.99$, $0.86$, and $0.76$, respectively.
For each problem instance (rows in \Cref{fig:optimization_landscapes}), four different properties are shown in the respective slice of the parameter space (columns in \Cref{fig:optimization_landscapes}): 
\begin{inparaenum}[(i)]
    \item the expectation value for VQE as per \Cref{eq:expectation},
    \item the estimated fidelity for ACAE as per \Cref{eq:fidelity} of the respective quantum state prepared by the Ansatz to the approximate eigenvector $\vec{v}_\text{appr.}$,
    \item the actual fidelity of the quantum state to $\vec{v}_\text{appr.}$, and
    \item the actual fidelity of the quantum state to an optimal reference eigenvector $\vec{v}_\text{opt.}$.
\end{inparaenum}
As in the experiments above, the VQE expectation values were computed from $200$ shots per circuit, whereas the fidelity was estimated based on $400$ snapshots.
The circuits were executed on a noiseless simulator, actual fidelities were determined based on state vector simulation provided by Qiskit.\looseness=-1

Due to the construction of WS-VQE, the parameters are optimized for the estimated fidelity of ACAE (second column) first, before the optimization continues for the expectation value of VQE (first column).
Therefore, it is essential for WS-VQE that a high fidelity is observed in proximity to low expectation values.
\Cref{fig:optimization_landscapes} illustrates a proximity of the respective extrema of both properties in the parameter space, which depends on the quality of the approximation.
For example, the minimal expectation value for problem 1 appears at parameter values very close to those of the maximum estimated and actual fidelity to the approximate eigenvector.
Since the approximate eigenvector for problems 2 and 3 are of lower quality, the similarities appear to fade progressively.
Meanwhile, the actual fidelity to an optimal eigenvector (rightmost column) appears to still exhibit its maximum at a similar location for problem 2.
However, since only one single optimal eigenvector is considered in the illustration of the landscapes, 
the extrema of the fidelity do not necessarily coincide with those of VQE, 
as is the case for problem 3.
We assume that there are other optimal or near optimal eigenvectors that make the minima of the expectation value appear in a different position.
In other words, the quantum state of the fidelity maximum for the specific optimal eigenvector of problem 3 may have a less significant influence on the expectation value than other solutions.
This is also consistent with the state fidelity to this eigenvector reaching a maximum value of only $0.22$, as compared to higher values for the other problems.\looseness=-1

One particularly noteworthy observation in \Cref{fig:optimization_landscapes} is that all fidelities have only one clearly pronounced maximum in the respective two-dimensional space (for the estimated fidelity: an area where the highest values are concentrated).
In contrast, the expectation values for all problem instances exhibit local minima.
These observations take into account the periodicity of the analyzed parameter space with a period of $2\pi$ due to the simple rotation gates of the ansatz.
In conjunction with a proximity of fidelity maxima for good approximate eigenvectors and expectation minima, this indicates that WS-VQE might be able to avoid local optima due to initial parameter optimization based on the estimated fidelity.

\section{Discussion}
\label{sec:dis}
The evaluation shows that WS-VQE based on ACAE can be worthwhile, but a few caveats remain to be discussed:\looseness=-1
\subsubsection{Quality of the Approximation}
Clearly, the success of WS-VQE depends on the quality of the approximations fed to the algorithm.
On average, the approximations corresponded to an approximation ratio around $0.850$ ($\sigma=0.126$) in our evaluation in \Cref{sec:results}.
We utilized a simple, but not generally feasible approximation for eigenvectors, as the inverse power method requires matrix inversion or solving linear systems of equations, which incur a cost of cubic complexity similar to that of the eigenvalue problem itself~\cite{pan1999complexity}.
Therefore, it would be desirable to utilize more efficient approximations.
These could be obtained, e.g., from previous solutions of periodically recurring problems with only little changes to the matrix, which are well-conditioned according to the Bauer-Fike theorem~\cite{bauer1960norms,quarteroni2006numerical}.
We emphasize that our work is focussed on the warm-starting technique behind WS-VQE itself and does not intend to prove a quantum advantage.\looseness=-1

\subsubsection{Performance of ACAE}
Another important factor in WS-VQE is the performance of ACAE.
In our evaluation, we observed an average estimated fidelity of $0.625$ ($\sigma=0.230$) achieved within the $50$ ACAE iterations.
Although the approximations obtained from the inverse power method were of relatively high quality, the low average fidelity indicates that ACAE was hardly able to retain this quality in its encoding.
Despite the seemingly low fidelity, WS-VQE still surpassed the performance of standard VQE.
However, another allocation of quantum resources for ACAE and WS-VQE, respectively, in particular providing more resources for the optimization of ACAE, could yield a higher fidelity and, thus, better results.

\subsubsection{Expressivity of the Ansatz}
An important prerequisite of the VQE is an Ansatz that is tunable to prepare the desired optimal solution.
This translates to the requirements of expressivity and trainability, i.e., the Ansatz needs to be expressive enough to capture the optimal solution, but simple enough that its parameter space can be traversed by the optimizer to reach an optimum~\cite{funcke2021dimensional,Du2022Expressivity}.
Likewise, the ACAE requires an expressive and trainable Ansatz for the encoding of complex amplitudes.
Although the Ansätze for the VQE and ACAE in principle need not be identical, WS-VQE implicitly assumes that an Ansatz for the VQE is also suitable to encode an approximate eigenvector with ACAE.
This assumption appears reasonable, particularly since ACAE only aims for an approximate encoding.
Provided that the VQE Ansatz is expressive, i.e., able to capture the optimum, and trainable, which entails the ability to reach states \enquote{surrounding} the optimum during the optimization toward it, and that the approximate solution $\vec{v}$ is sufficiently close to the optimal solution, the same Ansatz should also suffice to approximately encode $\vec{v}$.
\looseness=-1

\subsubsection{Choice of Classical Optimizers}
The choice of the classical optimizer in VQE has significant impact on the efficiency and performance, as it directly affects the number of measurements and iterations required and can mitigate adverse effects such as barren plateaus~\cite{Tilly2022VQEreview}.
Particularly, gradient-free optimizers require significantly fewer function evaluations, as they omit evaluating points in the parameter space to compute gradients.
For the evaluation, we used the same gradient-free optimizer for both ACAE and (WS-)VQE.
However, it may be more beneficial to select different optimizers for each optimization, since the optimization landscapes of both algorithms may exhibit different properties, as was illustrated in \Cref{sec:landscapes}.
Moreover, some optimizers are known to be more resilient to the noise on quantum devices~\cite{pellow2021comparison}.
Therefore, different optimizers may be required when executing \mbox{(WS-)VQE} on noisy quantum devices.

\subsubsection{Mitigation of Adverse Effects}
Parameter initializations have been shown to mitigate adverse effects in the optimization of VQAs~\cite{grant2019initialization,Lee2021parameterFixing}.
Particularly, barren plateaus, i.e., areas with vanishing gradients in the parameter space of the cost function, and local minima can be avoided with a viable parameter initialization.
Subsequently, the convergence of VQAs may improve significantly when the optimization is started closer to a global optimum.
We assume that the benefits of WS-VQE are partially due to the avoidance of the adverse effects by means of a viable parameter initialization, which is supported by our analysis in \Cref{sec:landscapes}. 
However, additional evaluation is needed to quantify their mitigation.
\looseness=-1

\subsubsection{VQE and Classical Shadows} 
As mentioned in~\Cref{sec:rw}, the idea of implementing the VQE based on classical shadows has emerged.
WS-VQE also combines the VQE with classical shadows, albeit differently.
Additionally, WS-VQE is fully compatible and complementary to a VQE implementation with classical shadows.
Recall that ACAE requires only an estimated fidelity, that is obtained with classical shadows from relatively few random Clifford measurements.
In contrast, the VQE may still require a multitude of measurements even when classical shadows are exploited, depending on the Pauli composition of the operator~\cite{Tilly2022VQEreview}.
Hence, WS-VQE with ACAE pretraining could still be more cost-efficient than starting a VQE implementation with classical shadows from scratch.

\subsubsection{Reuse of Clifford Unitaries}
As mentioned in \Cref{subsec:shots}, we reduced the classical effort of the Clifford measurements in ACAE by reusing the unitaries for the classical snapshots throughout an optimization.
Using the same unitaries for fidelity estimations throughout the optimization could lead to overfitting, i.e., the model could learn to accommodate the selected set of unitaries instead of actually improving the state fidelity. 
On the other hand, changing the unitaries in each iteration may increase the perturbation of the fidelity estimation, making the optimization more vulnerable.
However, we used relatively many snapshots, which mitigates both potential problems to a certain extent.
Moreover, our results indicate that the fidelity and encoding achieved by ACAE was sufficient to retain WS-VQE's performance benefit.
As an alternative, the random Clifford unitaries could be sampled from a reduced set that effectively preserves the original quantum channel introduced in \Cref{subsec:ACAE}, thus reducing the sampling and compilation costs~\cite{zhang2023minimal}.

\subsubsection{Number of Iterations and Queuing Times}
Despite potentially reducing the total number of shots needed, WS-VQE increases the total number of iterations due to the pretraining.
Since quantum computers are often available as shared resources through cloud services~\cite{Leymann2020_QuantumCloud,Vietz2021_QuantumSoftwareEngineeringChallenges}, waiting times in queues for the execution of quantum circuits could increase with the number of iterations.
Appropriate execution patterns supported by quantum cloud service providers mitigate this drawback by enabling a prioritized execution of VQAs, thus, removing the need to wait in queues for every iteration~\cite{Georg2023_PatternsQuantumExecution}.
\looseness=-1

\section{Summary and Outlook}
\label{sec:summary}
In this work, we proposed a warm-starting technique for the VQE that utilizes classical shadows.
In particular, the warm-start is based on an approximate amplitude encoding of an approximate solution for the problem at hand.
The proposed technique enables utilizing any given approximation of the problem that determines an (approximate) eigenvector to generate a beneficial parameter initialization for the VQE.
WS-VQE is also compatible with different variants and improvements of the VQE, as it does not require any changes of the original algorithm or its quantum circuit.
As shown in the evaluation, the VQE benefits from the warm-start in terms of a reduced quantum computational effort needed to reach a certain solution quality.
In addition, the evaluation showed that adjusting the optimizer used in WS-VQE can further improve the performance.
As an auxiliary result, we outlined a way to derive a parameter initialization from a biased initial state, which may serve as a blueprint for warm-starting other VQAs.
\looseness=-1

For future work, it remains to evaluate WS-VQE in more detail and with real-world use cases to determine which applications could benefit from it.
As explained in~\Cref{sec:dis}, changes to the configuration and implementation could improve the warm-starting technique further.
Moreover, we aim to analyze which other VQAs could benefit from this technique or adaptations of it.
Furthermore, our Quokka ecosystem~\cite{Beisel2023_Quokka} could be extended for the integration of WS-VQE-like warm-starts.
The Quokka ecosystem's \textit{warm-starting service} currently facilitates the \textit{classical} precomputation of parameter initializations and biased initial states, which could be extended for the hybrid quantum-classical pretraining of WS-VQE.

\section*{Acknowledgment}
This work was partially funded by the BMWK projects \textit{EniQmA} (01MQ22007B)
and \textit{SeQuenC} (01MQ22009B).

\footnotesize
\bibliography{Bib/bibliography,Bib/MasterBibliographyReduced}
\bibliographystyle{IEEEtranN}
\vspace{12pt}

\end{document}

%% file: Content/abstract.tex
The Variational Quantum Eigensolver (VQE) is a Variational Quantum Algorithm (VQA) to determine the ground state of quantum-mechanical systems.
As a VQA, it makes use of a classical computer to optimize parameter values for its quantum circuit.
However, each iteration of the VQE requires a multitude of measurements, and the optimization is subject to obstructions, such as barren plateaus, local minima, and subsequently slow convergence.
We propose a warm-starting technique, that utilizes an approximation to generate beneficial initial parameter values for the VQE aiming to mitigate these effects.
The warm-start is based on Approximate Complex Amplitude Encoding, a VQA using fidelity estimations from classical shadows to encode complex amplitude vectors into quantum states.
Such warm-starts open the path to fruitful combinations of classical approximation algorithms and quantum algorithms.
In particular, the evaluation of our approach shows that the warm-started VQE reaches higher quality solutions earlier than the original VQE.

%% file: Figures/circuit_subscripts.tex
\Qcircuit @C=1.0em @R=0.2em @!R { \\
	 	\nghost{{q}_{0} :  } & \lstick{{q}_{0} :  } & \gate{\mathrm{R_Y}\,(\mathrm{{\ensuremath{\theta}}_{0}})} & \gate{\mathrm{R_Z}\,(\mathrm{{\ensuremath{\theta}}_{3}})} \barrier[0em]{2} & \qw & \qw & \ctrl{1} \barrier[0em]{2} & \qw & \gate{\mathrm{R_Y}\,(\mathrm{{\ensuremath{\theta}}_{6}})} & \gate{\mathrm{R_Z}\,(\mathrm{{\ensuremath{\theta}}_{9}})} \barrier[0em]{2} & \qw & \qw & \ctrl{1} \barrier[0em]{2} & \qw & \gate{\mathrm{R_Y}\,(\mathrm{{\ensuremath{\theta}}_{12}})} & \gate{\mathrm{R_Z}\,(\mathrm{{\ensuremath{\theta}}_{15}})} & \qw & \qw\\
	 	\nghost{{q}_{1} :  } & \lstick{{q}_{1} :  } & \gate{\mathrm{R_Y}\,(\mathrm{{\ensuremath{\theta}}_{1}})} & \gate{\mathrm{R_Z}\,(\mathrm{{\ensuremath{\theta}}_{4}})} & \qw & \ctrl{1} & \targ & \qw & \gate{\mathrm{R_Y}\,(\mathrm{{\ensuremath{\theta}}_{7}})} & \gate{\mathrm{R_Z}\,(\mathrm{{\ensuremath{\theta}}_{10}})} & \qw & \ctrl{1} & \targ & \qw & \gate{\mathrm{R_Y}\,(\mathrm{{\ensuremath{\theta}}_{13}})} & \gate{\mathrm{R_Z}\,(\mathrm{{\ensuremath{\theta}}_{16}})} & \qw & \qw\\
	 	\nghost{{q}_{2} :  } & \lstick{{q}_{2} :  } & \gate{\mathrm{R_Y}\,(\mathrm{{\ensuremath{\theta}}_{2}})} & \gate{\mathrm{R_Z}\,(\mathrm{{\ensuremath{\theta}}_{5}})} & \qw & \targ & \qw & \qw & \gate{\mathrm{R_Y}\,(\mathrm{{\ensuremath{\theta}}_{8}})} & \gate{\mathrm{R_Z}\,(\mathrm{{\ensuremath{\theta}}_{11}})} & \qw & \targ & \qw & \qw & \gate{\mathrm{R_Y}\,(\mathrm{{\ensuremath{\theta}}_{14}})} & \gate{\mathrm{R_Z}\,(\mathrm{{\ensuremath{\theta}}_{17}})} & \qw & \qw\\
}

%% file: paper.bbl
\begin{thebibliography}{55}
\providecommand{\natexlab}[1]{#1}
\providecommand{\url}[1]{#1}
\csname url@samestyle\endcsname
\providecommand{\newblock}{\relax}
\providecommand{\bibinfo}[2]{#2}
\providecommand{\BIBentrySTDinterwordspacing}{\spaceskip=0pt\relax}
\providecommand{\BIBentryALTinterwordstretchfactor}{4}
\providecommand{\BIBentryALTinterwordspacing}{\spaceskip=\fontdimen2\font plus
\BIBentryALTinterwordstretchfactor\fontdimen3\font minus
  \fontdimen4\font\relax}
\providecommand{\BIBforeignlanguage}[2]{{%
\expandafter\ifx\csname l@#1\endcsname\relax
\typeout{** WARNING: IEEEtranN.bst: No hyphenation pattern has been}%
\typeout{** loaded for the language `#1'. Using the pattern for}%
\typeout{** the default language instead.}%
\else
\language=\csname l@#1\endcsname
\fi
#2}}
\providecommand{\BIBdecl}{\relax}
\BIBdecl

\bibitem[Feynman(1982)]{feynman1982simulating}
R.~P. Feynman, ``Simulating physics with computers,'' \emph{International
  Journal of Theoretical Physics}, vol.~21, no. 6/7, 1982.

\bibitem[Preskill(2018)]{preskill2018quantum}
J.~Preskill, ``Quantum computing in the {NISQ} era and beyond,''
  \emph{Quantum}, vol.~2, 2018.

\bibitem[Tilly et~al.(2022)Tilly, Chen, Cao, Picozzi, Setia, Li, Grant,
  Wossnig, Rungger, Booth, and Tennyson]{Tilly2022VQEreview}
J.~Tilly \emph{et~al.}, ``The variational quantum eigensolver: A review of
  methods and best practices,'' \emph{Physics Reports}, vol. 986, 2022.

\bibitem[Peruzzo et~al.(2014)Peruzzo, McClean, Shadbolt, Yung, Zhou, Love,
  Aspuru-Guzik, and O’brien]{peruzzo2014variational}
A.~Peruzzo \emph{et~al.}, ``A variational eigenvalue solver on a photonic
  quantum processor,'' \emph{Nature communications}, vol.~5, no.~1, 2014.

\bibitem[Leymann and Barzen(2020)]{Leymann2020_QuantumAlgorithmsBitterTurth}
\BIBentryALTinterwordspacing
F.~Leymann and J.~Barzen, ``{The bitter truth about gate-based quantum
  algorithms in the NISQ era},'' \emph{Quantum Science and Technology}, pp.
  1--28, Sep. 2020. [Online]. Available:
  \url{https://doi.org/10.1088/2058-9565/abae7d}
\BIBentrySTDinterwordspacing

\bibitem[Truger et~al.(2023)Truger, Barzen, Bechtold, Beisel, Leymann, Mandl,
  and Yussupov]{Truger2023_warm_starting}
F.~Truger \emph{et~al.}, ``{Warm-Starting and Quantum Computing: A Systematic
  Mapping Study},'' \emph{arXiv:2303.06133}, 2023.

\bibitem[Egger et~al.(2021)Egger, Mare{\v{c}}ek, and Woerner]{egger2021warm}
D.~J. Egger, J.~Mare{\v{c}}ek, and S.~Woerner, ``Warm-starting quantum
  optimization,'' \emph{Quantum}, vol.~5, 2021.

\bibitem[Tate et~al.(2023)Tate, Farhadi, Herold, Mohler, and
  Gupta]{tate2023bridging}
R.~Tate, M.~Farhadi, C.~Herold, G.~Mohler, and S.~Gupta, ``Bridging classical
  and quantum with sdp initialized warm-starts for qaoa,'' \emph{ACM
  Transactions on Quantum Computing}, vol.~4, no.~2, 2023.

\bibitem[Galda et~al.(2021)Galda, Liu, Lykov, Alexeev, and
  Safro]{galda2021transferability}
A.~Galda, X.~Liu, D.~Lykov, Y.~Alexeev, and I.~Safro, ``Transferability of
  optimal qaoa parameters between random graphs,'' in \emph{2021 IEEE
  International Conference on Quantum Computing and Engineering (QCE)}.\hskip
  1em plus 0.5em minus 0.4em\relax IEEE, 2021.

\bibitem[Mitarai et~al.(2022)Mitarai, Suzuki, Mizukami, Nakagawa, and
  Fujii]{mitarai2022quadratic}
K.~Mitarai, Y.~Suzuki, W.~Mizukami, Y.~O. Nakagawa, and K.~Fujii, ``Quadratic
  clifford expansion for efficient benchmarking and initialization of
  variational quantum algorithms,'' \emph{Physical Review Research}, vol.~4,
  no.~3, 2022.

\bibitem[Sack and Serbyn(2021)]{sack2021quantum}
S.~H. Sack and M.~Serbyn, ``Quantum annealing initialization of the quantum
  approximate optimization algorithm,'' \emph{quantum}, vol.~5, 2021.

\bibitem[Shaydulin et~al.(2023)Shaydulin, Lotshaw, Larson, Ostrowski, and
  Humble]{shaydulin2023parameter}
R.~Shaydulin, P.~C. Lotshaw, J.~Larson, J.~Ostrowski, and T.~S. Humble,
  ``Parameter transfer for quantum approximate optimization of weighted
  maxcut,'' \emph{ACM Transactions on Quantum Computing}, vol.~4, no.~3, 2023.

\bibitem[Mitsuda et~al.(2022)Mitsuda, Nakaji, Suzuki, Tanaka, Raymond, Tezuka,
  Onodera, and Yamamoto]{mitsuda2022ACAE}
N.~Mitsuda \emph{et~al.}, ``Approximate complex amplitude encoding algorithm
  and its application to data classification problems,''
  \emph{arXiv:2211.13039}, 2022.

\bibitem[Huang et~al.(2020)Huang, Kueng, and Preskill]{huang2020predicting}
H.-Y. Huang, R.~Kueng, and J.~Preskill, ``Predicting many properties of a
  quantum system from very few measurements,'' \emph{Nature Physics}, vol.~16,
  no.~10, 2020.

\bibitem[Cerezo et~al.(2021)Cerezo, Arrasmith, Babbush, Benjamin, Endo, Fujii,
  McClean, Mitarai, Yuan, Cincio, et~al.]{cerezo2021variational}
M.~Cerezo \emph{et~al.}, ``Variational quantum algorithms,'' \emph{Nature
  Reviews Physics}, vol.~3, no.~9, 2021.

\bibitem[Li et~al.(2019)Li, Hu, Zhang, Song, and
  Yung]{Li2019vqeQuantumChemistry}
Y.~Li, J.~Hu, X.-M. Zhang, Z.~Song, and M.-H. Yung, ``Variational quantum
  simulation for quantum chemistry,'' \emph{Advanced Theory and Simulations},
  vol.~2, no.~4, 2019.

\bibitem[Head-Marsden et~al.(2021)Head-Marsden, Flick, Ciccarino, and
  Narang]{HeadMarsden2021quantumInfomration}
K.~Head-Marsden, J.~Flick, C.~J. Ciccarino, and P.~Narang, ``Quantum
  information and algorithms for correlated quantum matter,'' \emph{Chemical
  Reviews}, vol. 121, no.~5, 2021.

\bibitem[Mustafa et~al.(2022)Mustafa, Morapakula, Jain, and
  Ganguly]{Mustafa2022drugDiscovery}
H.~Mustafa, S.~N. Morapakula, P.~Jain, and S.~Ganguly, ``Variational quantum
  algorithms for chemical simulation and drug discovery,'' in \emph{2022
  International Conference on Trends in Quantum Computing and Emerging Business
  Technologies (TQCEBT)}, 2022.

\bibitem[Barkoutsos et~al.(2021)Barkoutsos, Gkritsis, Ollitrault, Sokolov,
  Woerner, and Tavernelli]{barkoutsos2021quantum}
P.~K. Barkoutsos \emph{et~al.}, ``Quantum algorithm for alchemical optimization
  in material design,'' \emph{Chemical science}, vol.~12, no.~12, 2021.

\bibitem[Bernal et~al.(2022)Bernal, Ajagekar, Harwood, Stober, Trenev, and
  You]{bernal2022perspectives}
D.~E. Bernal \emph{et~al.}, ``Perspectives of quantum computing for chemical
  engineering,'' \emph{AIChE Journal}, vol.~68, no.~6, 2022.

\bibitem[Weigold et~al.(2020)Weigold, Barzen, Leymann, and
  Salm]{Weigold2020_DataEncodingPatterns}
M.~Weigold, J.~Barzen, F.~Leymann, and M.~Salm, ``{Data Encoding Patterns For
  Quantum Algorithms},'' in \emph{Proceedings of the 27\textsuperscript{th}
  Conference on Pattern Languages of Programs (PLoP '20)}.\hskip 1em plus 0.5em
  minus 0.4em\relax HILLSIDE, 2020.

\bibitem[Schuld and Petruccione(2018)]{schuld2018supervised}
M.~Schuld and F.~Petruccione, \emph{Supervised learning with quantum
  computers}.\hskip 1em plus 0.5em minus 0.4em\relax Springer, 2018, vol.~17.

\bibitem[Shende et~al.(2006)Shende, Bullock, and Markov]{Shende2006Synthesis}
V.~Shende, S.~Bullock, and I.~Markov, ``Synthesis of quantum-logic circuits,''
  \emph{IEEE Transactions on Computer-Aided Design of Integrated Circuits and
  Systems}, vol.~25, no.~6, 2006.

\bibitem[Nakaji et~al.(2022)Nakaji, Uno, Suzuki, Raymond, Onodera, Tanaka,
  Tezuka, Mitsuda, and Yamamoto]{nakaji2022AAE}
K.~Nakaji \emph{et~al.}, ``Approximate amplitude encoding in shallow
  parameterized quantum circuits and its application to financial market
  indicators,'' \emph{Physical Review Research}, vol.~4, no.~2, 2022.

\bibitem[Zhang et~al.(2021)Zhang, Kyaw, Kottmann, Degroote, and
  Aspuru-Guzik]{Zhang2021adaptiveAnsatzConstruction}
Z.-J. Zhang, T.~H. Kyaw, J.~S. Kottmann, M.~Degroote, and A.~Aspuru-Guzik,
  ``Mutual information-assisted adaptive variational quantum eigensolver,''
  \emph{Quantum Science and Technology}, vol.~6, no.~3, 2021.

\bibitem[Grimsley et~al.(2023)Grimsley, Barron, Barnes, Economou, and
  Mayhall]{grimsley2023adaptive}
H.~R. Grimsley, G.~S. Barron, E.~Barnes, S.~E. Economou, and N.~J. Mayhall,
  ``Adaptive, problem-tailored variational quantum eigensolver mitigates rough
  parameter landscapes and barren plateaus,'' \emph{npj Quantum Information},
  vol.~9, no.~1, 2023.

\bibitem[Verdon et~al.(2019)Verdon, Broughton, McClean, Sung, Babbush, Jiang,
  Neven, and Mohseni]{verdon2019meta}
G.~Verdon \emph{et~al.}, ``Learning to learn with quantum neural networks via
  classical neural networks,'' \emph{arXiv:1907.05415}, 2019.

\bibitem[Wilson et~al.(2021)Wilson, Stromswold, Wudarski, Hadfield, Tubman, and
  Rieffel]{wilson2021meta}
M.~Wilson \emph{et~al.}, ``Optimizing quantum heuristics with meta-learning,''
  \emph{Quantum Machine Intelligence}, vol.~3, 2021.

\bibitem[Tao et~al.(2023)Tao, Wu, Xia, and Li]{tao2023laws}
Z.~Tao, J.~Wu, Q.~Xia, and Q.~Li, ``Laws: Look around and warm-start natural
  gradient descent for quantum neural networks,'' in \emph{2023 IEEE
  International Conference on Quantum Software (QSW)}.\hskip 1em plus 0.5em
  minus 0.4em\relax IEEE, 2023.

\bibitem[Dborin et~al.(2022)Dborin, Barratt, Wimalaweera, Wright, and
  Green]{dborin2022matrix}
J.~Dborin, F.~Barratt, V.~Wimalaweera, L.~Wright, and A.~G. Green, ``Matrix
  product state pre-training for quantum machine learning,'' \emph{Quantum
  Science and Technology}, vol.~7, no.~3, 2022.

\bibitem[Rudolph et~al.(2022)Rudolph, Miller, Motlagh, Chen, Acharya, and
  Perdomo-Ortiz]{rudolph2022synergy}
M.~S. Rudolph \emph{et~al.}, ``Synergy between quantum circuits and tensor
  networks: Short-cutting the race to practical quantum advantage,''
  \emph{arXiv:2208.13673}, 2022.

\bibitem[Ravi et~al.(2022)Ravi, Gokhale, Ding, Kirby, Smith, Baker, Love,
  Hoffmann, Brown, and Chong]{Ravi2022CAFQA}
G.~S. Ravi \emph{et~al.}, ``Cafqa: A classical simulation bootstrap for
  variational quantum algorithms,'' in \emph{Proceedings of the 28th ACM
  International Conference on Architectural Support for Programming Languages
  and Operating Systems, Volume 1}, ser. ASPLOS 2023.\hskip 1em plus 0.5em
  minus 0.4em\relax Association for Computing Machinery, 2022.

\bibitem[Cervera-Lierta et~al.(2021)Cervera-Lierta, Kottmann, and
  Aspuru-Guzik]{cervera2021meta}
A.~Cervera-Lierta, J.~S. Kottmann, and A.~Aspuru-Guzik, ``Meta-variational
  quantum eigensolver: Learning energy profiles of parameterized hamiltonians
  for quantum simulation,'' \emph{PRX Quantum}, vol.~2, no.~2, 2021.

\bibitem[Harwood et~al.(2022)Harwood, Trenev, Stober, Barkoutsos, Gujarati,
  Mostame, and Greenberg]{Harwood2022VAQCVQE}
S.~M. Harwood \emph{et~al.}, ``Improving the variational quantum eigensolver
  using variational adiabatic quantum computing,'' \emph{ACM Transactions on
  Quantum Computing}, vol.~3, no.~1, 2022.

\bibitem[Skogh et~al.(2023)Skogh, Leinonen, Lolur, and
  Rahm]{skogh2023accelerating}
M.~Skogh, O.~Leinonen, P.~Lolur, and M.~Rahm, ``Accelerating variational
  quantum eigensolver convergence using parameter transfer,'' \emph{Electronic
  Structure}, vol.~5, no.~3, 2023.

\bibitem[Kanno and Tada(2021)]{kanno2021many}
S.~Kanno and T.~Tada, ``Many-body calculations for periodic materials via
  restricted boltzmann machine-based vqe,'' \emph{Quantum Science and
  Technology}, vol.~6, no.~2, 2021.

\bibitem[Beisel et~al.(2023)Beisel, Barzen, Garhofer, Leymann, Truger, Weder,
  and Yussupov]{Beisel2023_Quokka}
M.~Beisel \emph{et~al.}, ``{Quokka: A Service Ecosystem for Workflow-Based
  Execution of Variational Quantum Algorithms},'' in \emph{Service-Oriented
  Computing -- ICSOC 2022 Workshops}.\hskip 1em plus 0.5em minus 0.4em\relax
  Springer, 2023, Demonstration.

\bibitem[Huang et~al.(2021)Huang, Kueng, and Preskill]{huang2021derandomizedCS}
H.-Y. Huang, R.~Kueng, and J.~Preskill, ``Efficient estimation of pauli
  observables by derandomization,'' \emph{Phys. Rev. Lett.}, vol. 127, 2021.

\bibitem[{Qiskit contributors}(2023)]{qiskit}
{Qiskit contributors}, ``Qiskit: An open-source framework for quantum
  computing,'' 2023.

\bibitem[Grant et~al.(2019)Grant, Wossnig, Ostaszewski, and
  Benedetti]{grant2019initialization}
E.~Grant, L.~Wossnig, M.~Ostaszewski, and M.~Benedetti, ``An initialization
  strategy for addressing barren plateaus in parametrized quantum circuits,''
  \emph{Quantum}, vol.~3, 2019.

\bibitem[Quarteroni et~al.(2006)Quarteroni, Sacco, and
  Saleri]{quarteroni2006numerical}
A.~Quarteroni, R.~Sacco, and F.~Saleri, \emph{Numerical mathematics}.\hskip 1em
  plus 0.5em minus 0.4em\relax Springer Science \& Business Media, 2006,
  vol.~37.

\bibitem[Gershgorin(1931)]{gershgorin1931circles}
S.~Gershgorin, ``\BIBforeignlanguage{German}{{\"U}ber die {Abgrenzung} der
  {Eigenwerte} einer {Matrix}.}'' \emph{\BIBforeignlanguage{German}{Bull. Acad.
  Sci. USSR, ser. VII}}, no.~6, 1931.

\bibitem[Funcke et~al.(2021)Funcke, Hartung, Jansen, K{\"u}hn, and
  Stornati]{funcke2021dimensional}
L.~Funcke, T.~Hartung, K.~Jansen, S.~K{\"u}hn, and P.~Stornati, ``Dimensional
  expressivity analysis of parametric quantum circuits,'' \emph{Quantum},
  vol.~5, 2021.

\bibitem[Bravyi and Maslov(2021)]{Bravyi2021Clifford}
S.~Bravyi and D.~Maslov, ``Hadamard-free circuits expose the structure of the
  clifford group,'' \emph{IEEE Transactions on Information Theory}, vol.~67,
  no.~7, 2021.

\bibitem[Powell(1994)]{cobyla}
M.~J.~D. Powell, \emph{A Direct Search Optimization Method That Models the
  Objective and Constraint Functions by Linear Interpolation}.\hskip 1em plus
  0.5em minus 0.4em\relax Springer Netherlands, 1994.

\bibitem[Pellow-Jarman et~al.(2021)Pellow-Jarman, Sinayskiy, Pillay, and
  Petruccione]{pellow2021comparison}
A.~Pellow-Jarman, I.~Sinayskiy, A.~Pillay, and F.~Petruccione, ``A comparison
  of various classical optimizers for a variational quantum linear solver,''
  \emph{Quantum Information Processing}, vol.~20, no.~6, 2021.

\bibitem[Truger and Obst(2024)]{gitRepo}
\BIBentryALTinterwordspacing
F.~Truger and J.~Obst, ``{WS-VQE-prototype GitHub Repository},'' 2024.
  [Online]. Available: \url{https://github.com/UST-QuAntiL/WS-VQE-prototype}
\BIBentrySTDinterwordspacing

\bibitem[Pan and Chen(1999)]{pan1999complexity}
V.~Y. Pan and Z.~Q. Chen, ``The complexity of the matrix eigenproblem,'' in
  \emph{Proceedings of the thirty-first annual ACM symposium on Theory of
  computing}, 1999.

\bibitem[Bauer and Fike(1960)]{bauer1960norms}
F.~L. Bauer and C.~T. Fike, ``Norms and exclusion theorems,'' \emph{Numerische
  Mathematik}, vol.~2, 1960.

\bibitem[Du et~al.(2022)Du, Tu, Yuan, and Tao]{Du2022Expressivity}
Y.~Du, Z.~Tu, X.~Yuan, and D.~Tao, ``Efficient measure for the expressivity of
  variational quantum algorithms,'' \emph{Phys. Rev. Lett.}, vol. 128, 2022.

\bibitem[Lee et~al.(2021)Lee, Saito, Cai, and Asai]{Lee2021parameterFixing}
X.~Lee, Y.~Saito, D.~Cai, and N.~Asai, ``Parameters fixing strategy for quantum
  approximate optimization algorithm,'' in \emph{IEEE International Conference
  on Quantum Computing and Engineering (QCE)}, 2021.

\bibitem[Zhang et~al.(2023)Zhang, Liu, and Zhou]{zhang2023minimal}
Q.~Zhang, Q.~Liu, and Y.~Zhou, ``Minimal clifford shadow estimation by mutually
  unbiased bases,'' \emph{arXiv:2310.18749}, 2023.

\bibitem[Leymann et~al.(2020)Leymann, Barzen, Falkenthal, Vietz, Weder, and
  Wild]{Leymann2020_QuantumCloud}
F.~Leymann \emph{et~al.}, ``{Quantum in the Cloud: Application Potentials and
  Research Opportunities},'' in \emph{Proceedings of the 10\textsuperscript{th}
  International Conference on Cloud Computing and Services Science (CLOSER
  2020)}.\hskip 1em plus 0.5em minus 0.4em\relax SciTePress, 2020.

\bibitem[Vietz et~al.(2021)Vietz, Barzen, Leymann, Weder, and
  Yussupov]{Vietz2021_QuantumSoftwareEngineeringChallenges}
D.~Vietz, J.~Barzen, F.~Leymann, B.~Weder, and V.~Yussupov, ``{An Exploratory
  Study on the Challenges of Engineering Quantum Applications in the Cloud},''
  in \emph{Proceedings of the 2\textsuperscript{nd} Quantum Software
  Engineering and Technology Workshop (Q-SET 2021) co-located with IEEE
  International Conference on Quantum Computing and Engineering (QCE21)}.\hskip
  1em plus 0.5em minus 0.4em\relax CEUR Workshop Proceedings, 2021, workshop.

\bibitem[Georg et~al.(2023)Georg, Barzen, Beisel, Leymann, Obst, Vietz, Weder,
  and Yussupov]{Georg2023_PatternsQuantumExecution}
D.~Georg \emph{et~al.}, ``{Execution Patterns for Quantum Applications},'' in
  \emph{Proceedings of the 18\textsuperscript{th} International Conference on
  Software Technologies - ICSOFT}.\hskip 1em plus 0.5em minus 0.4em\relax
  SciTePress, 2023.

\end{thebibliography}
